\documentclass[%
 reprint,
superscriptaddress,
nofootinbib,
 amsmath,amssymb,
 aps,
]{revtex4-2}
\usepackage{physics}
\usepackage{graphicx}% Include figure files
\usepackage{dcolumn}% Align table columns on decimal point
\usepackage{bm}% bold math
\usepackage{hyperref}% add hypertext capabilities
\usepackage[USenglish]{babel}
\usepackage{comment}

\begin{document}

\preprint{APS/123-QED}

\title{Wigner-Seitz truncated TDDFT approach for the \\ calculation of exciton binding energies in solids}% Force line breaks with \\

\author{M. Arruabarrena}%
\affiliation{%
 Centro de F\'isica de Materiales - Materials Physics Center (CFM-MPC), 20018 Donostia, Spain\\
}%
\author{A. Leonardo}
\affiliation{%
 Centro de F\'isica de Materiales - Materials Physics Center (CFM-MPC), 20018 Donostia, Spain\\
}%
\affiliation{%
 Donostia International Physics Center (DIPC), 20018 Donostia, Spain\\
}%
\affiliation{%
 EHU Quantum Center, Universidad del País Vasco/Euskal Herriko Unibertsitatea UPV/EHU, Leioa, Spain\\
}%
%\email{aritz.leonardo@ehu.eus}
\author{A. Ayuela}
\affiliation{%
 Centro de F\'isica de Materiales - Materials Physics Center (CFM-MPC), 20018 Donostia, Spain\\
}%
\affiliation{%
 Donostia International Physics Center (DIPC), 20018 Donostia, Spain\\
}%


\begin{abstract}
    Time Dependent Density Functional Theory (TDDFT) has been currently established as a computationally cheaper, yet effective, alternative to the Many Body Perturbation Theory (MBPT) for calculating the optical properties of solids. Within the Linear Response formalism, the optical absorption spectra are in good agreement with experiments, as well as the direct determination of the exciton binding energies. However, the family of exchange-correlation kernels known as long-range corrected (LRC) kernels  that correctly capture excitonic features have difficulties  for simultaneously producing good-looking spectra and accurate exciton binding energies. More recently, this discrepancy has been partially overcomed by a hybrid-TDDFT approach. We show that the key resides in the numerical treatment of the long-range Coulomb singular term. We carefully study the effect of this term, both in the pure-TDDFT and hybrid approach using a Wigner-Seitz truncated kernel. We find that computing this term presents technical difficulties that are hard to overcome in both approaches, and that points to the need for a better description of the electron-hole interaction. 
\end{abstract}

\maketitle

\section{Introduction} 
A challenging task of material science is the correct description of excitation phenomena. The design of novel man-made materials with desired properties relies upon the capability to accurately predict their electronic structure, but calculations for interacting electrons are much more challenging than those of independent electrons. Over the last two decades, TDDFT has proven to be very successful for the calculation of neutral excitations both for finite and extended systems and it is often considered as the computational cheaper but rigorous alternative to the state-of-the-art MBPT \cite{Onida2002, Botti_2007}. 

Excitonic effects in semiconductors and insulating materials have been successfully described by the Linear response formalism of TDDFT, solving a Dyson type equation to obtain the density response function which yields good looking optical spectra of several materials \cite{LRC-Botti,Sharma2011, Rigamonti2015}. On the other hand, a completely equivalent approach based on the same formalism reformulates the equations into an eigenvalue problem allowing a direct determination of exciton binding energies, without explicitly having to compute the response function \cite{Ullrich2014, Yang2013, Byun2017}. Originally for finite system, these equations were known as the Casida equations\cite{CASIDA1995} and were later extended to periodic solids \cite{Turkowski2009}. In the two methods, excitonic features are correctly captured using the family of exchange-correlation kernels known as long-range corrected (LRC) kernels; however, these kernels are not capable to produce simultaneously good looking spectra and accurate exciton binding energies. In other words, the material dependent parameter $\alpha$ of the LRC-type kernels fitted to provide good optical spectra of semiconductors, may differ by more than an order of magnitude, to the same parameter fitted to estimate experimental binding energies of excitons  \cite{Byun2017}. As it will be justified in the following sections, we believe that the above mentioned discrepancy has a numerical origin, as the main contribution towards exciton binding energies come from a singular term (\textbf{q}=0) of the kernel, which is ill-defined for periodic solids\cite{Resta-Position-operator, Gu2013}.\\

Lately, hybrid approaches between TDDFT and the Bethe-Salpeter Equation (BSE) equation have been proposed \cite{SXX-paper, Byun_Ullrich-kernels_2020} with success having an affordable computational cost. Furthermore, a hybrid exchange-correlation term  can be considered by mixing an adiabatic TDDFT kernel with screened exact exchange (SXX), so that good exciton binding energies were obtained in conjunction with optical spectra \cite{Sun-Yang-Ullrich-2020, Sun-Ullrich-perovskites}.  In fact the head-only SXX kernel including the divergent term \textbf{q=0} yielded identical binding energies to the full kernel \cite{Sun-Yang-Ullrich-2020}. These findings indicate that the singular term is still crucial also in this approach, and thus, should be studied in detail when dealing with exciton calculations. \\ 

In this work, we carefully study the effect of the singular long-range Coulomb term on the calculation of exciton binding energies. 
On the one hand, in the pure TDDFT framework, we report the absence of a correction term $\mathbf{C}_{c\mathbf{k},v\mathbf{k}}$ that has to be included when applying the commutator relation that is often used to deal with singularities \cite{Gu2013}. We compute the magnitude of this term for a set of semiconductors, and find that it is often larger than the regularized singular term itself, which could help explaining the discrepancy between the absorption spectra and the  calculations of exciton binding energies. On the other hand, in the hybrid framework, we propose a Wigner-Seitz Truncated SXX kernel (WS-SXX) with a well defined analytical term in the optical limit (\textbf{q}=0). We furthermore analyze the performance of the kernel with respect to the different convergence parameters, and compare it with the state-of-the art methods in the field.

%%%%%%%%%%%%%%%%%%%%%%%%%%%%%%%%%%%%%%%%%%%%%%%%%%%%%%%%%%%% 
%%%%%%%%%%%%%%%%%%%%%%%%%  THEORY  %%%%%%%%%%%%%%%%%%%%%%%%%
%%%%%%%%%%%%%%%%%%%%%%%%%%%%%%%%%%%%%%%%%%%%%%%%%%%%%%%%%%%% 
\section{THEORY: Exciton binding energies}
Both in TDDFT and within the many body BSE formalism, it is derived a formally equivalent eigenvalue problem to calculate directly the excitation energies of a given system \cite{CASIDA1995,martin_reining_ceperley_2016}:
%known as Casida equations 
\begin{equation}
%\mathsf{P} =
\begin{pmatrix}
\mathbf{A} & \mathbf{B}  \\
\mathbf{B}^* & \mathbf{A}^*  \\
\end{pmatrix}
\begin{pmatrix}
\textbf{X}_n  \\
\textbf{Y}_n \\
\end{pmatrix}
= \omega_n
\begin{pmatrix}
\mathbf{-1} & \mathbf{0}  \\
\mathbf{0} & \mathbf{1} \\
\end{pmatrix}
\begin{pmatrix}
\textbf{X}_n  \\
\textbf{Y}_n \\
\end{pmatrix}
\label{casida_eigen} 
\end{equation}
where $\mathbf{A}$ and $\mathbf{B}$ are excitation and de-excitation matrices, respectively, and \textbf{X}$_n$ and \textbf{Y}$_n$ are the $n$th eigenvectors of the $\omega_n$ excitation energy (the $n$-th eigenvalue). In both formalisms, the Tamm-Dancoff approximation (TDA), which decouples excitations and de-excitations, is widely used to reduce the dimension of the eigenvalue problem and is equivalent to setting \textbf{B} to zero. Equation (\ref{casida_eigen}) applied to periodic crystals is thus rewritten as  
\begin{equation}
    [(\epsilon_{c\textbf{k}}-\epsilon_{v\textbf{k}'})\delta_{vv'}\delta_{cc'}\delta_{\textbf{k}\textbf{k}'} + F_{cv\textbf{k},c'v'\textbf{k}'}^{\text{Hxc}}] \textbf{Y}_n = \omega_n \textbf{Y}_n. 
\end{equation}
Here, ($v \mathbf{k}$) and ($c \mathbf{k}$) indicate Bloch waves within the first Brillouin zone, where $v$ and $c$ are the valence and conduction indices, respectively. \\ 

In fact, the main difference between the TDDFT and BSE approaches lies in the coupling matrix $F^{Hxc} = F^H + F^{xc}$. 
The Hartree kernel is the same in both approaches given as: 
\begin{equation}
    F_{cv\textbf{k},c'v'\textbf{k}'}^{\text{H}} = \frac{2}{V} \sum_{\textbf{G}\ne 0} \frac{4\pi}{\abs{\textbf{G}}^2} \bra{c\textbf{k}}e^{i\textbf{G}\cdot\textbf{r}}\ket{v\textbf{k}}\bra{v'\textbf{k}'}e^{-i\textbf{G}\cdot\textbf{r}}\ket{c'\textbf{k}'}. 
\end{equation}
%The long range part of the Hartree term is omitted so that the eigenvalues of Eq.~(\ref{casida_eigen}) correspond to the poles of the dielectric function. 
The exchange-correlation F$^{xc}$ term of the coupling matrix differs for the TDDFT and BSE approaches. 
Within the optical limit, since the momentum transfer in negligible $(\mathbf{q}\rightarrow 0)$, the matrix elements in TDDFT are given by 
\begin{align}
    F_{cv\textbf{k},c'v'\textbf{k}'}^{xc} &= \frac{2}{V} \sum_{\textbf{G},\textbf{G}'} f_{xc,\textbf{GG}'}(\textbf{q}\rightarrow 0) \nonumber \\ 
    &\times\bra{c\textbf{k}}e^{i(\textbf{q} + \textbf{G})\cdot\textbf{r}}\ket{v\textbf{k}}\bra{v'\textbf{k}'}e^{-i(\textbf{q}+\textbf{G}')\cdot\textbf{r}}\ket{c'\textbf{k}'}, 
    \label{eq:xc-tddft}
\end{align} 
where $f_{xc}$ is the exchange-correlation kernel, and $V$ is the crystal volume. In the BSE formalism, the term has the form 
\begin{align}
    F^{\text{xc}}_{cv\textbf{k},c'v'\textbf{k}'} &= -\frac{1}{V} \sum_{\textbf{G,G'}}  W_{\textbf{G},\textbf{G}'}(\textbf{q},\omega) \delta_{\textbf{q},\textbf{k}-\textbf{k}'} \nonumber \\
    &\times \bra{c\textbf{k}}e^{i(\textbf{q} + \textbf{G})\cdot\textbf{r}}\ket{c'\textbf{k}'}\bra{v'\textbf{k}'}e^{-i(\textbf{q}+\textbf{G}')\cdot\textbf{r}}\ket{v\textbf{k}}, 
    \label{eq:xc-BSE}
\end{align}
where $W_{\textbf{G},\textbf{G}'}$ is the screened Coulomb interaction. Note that in TDDFT the xc matrix elements involve braket operations between valence and conduction bands, and in MBPT these elements  are given by $c$-$c'$ and $v$-$v'$ integrals.

\section{RESULTS}
\subsection{TDDFT approach} 
\begin{figure*}[hpt]
    \centering
    \includegraphics[scale=0.75]{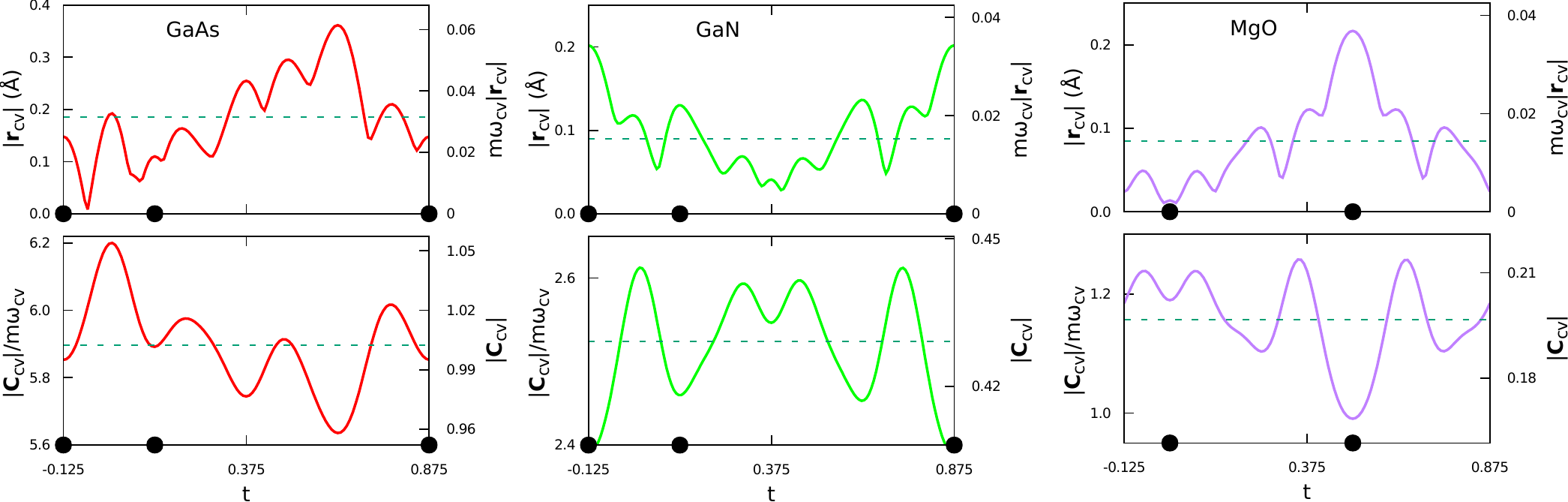}
    \caption{Magnitude of the r-matrix (top) and C-matrix (bottom) elements in bulk GaAs, %$\beta$-GaN 
    z-GaN and MgO. The magnitudes in the left axes are given in $\text{\AA}$, and those in the right axes are in units of the momentum matrix element $p_{cv}$. The horizontal axis represents the cell center on the [111] direction of the conventional unit cell, so that the positions of the cell center are given by $t(a,a,a)$. The black dots in these axes indicate the positions of the atoms, and the dashed horizontal lines represent the mean values of the plotted quantities.   }
    \label{fig:2}
\end{figure*} 
Since TDDFT is a formally rigorous theory, the solution of eq. (\ref{eq:xc-tddft}) should give the exact values of the excitonic energies, as long as we are able to satisfactorily approximate the unknown exchange-correlation kernel. The family of exchange-correlation kernels known as long-range corrected (LRC) kernels have shown to correctly capture excitonic features by reproducing experimental absorption spectra in semiconductors \cite{Sharma2011, LRC-Botti, Byun2017, PhysRevLett.114.146402,Byun_Ullrich-kernels_2020}. The agreement with experimental absorption spectra is remarkable even when we employ its simplest static form of
\begin{align}
f_{xc}^{\text{LRC}} = \frac{- \alpha\,\delta_{\textbf{GG}'}}{\abs{\mathbf{q+G}}^2}.
\end{align} %\\ 
However, the empirical parameter $\alpha$ determined from the static dielectric constant of the crystal $(\alpha=4.615 \varepsilon_{\infty}^{-1}-0.213)$ fails to reproduce bound excitons, unless $\alpha$ is
 set ad-hoc to a much higher value (see Fig. [1] in Ref. \cite{Byun2017}). A possible solution to the discrepancy is to propose a material-dependent non-uniform scaling factor for the Bootstrap-type kernels, so that the peak height and the position are correctly reproduced. The price to pay is including an extra arbitrary function without theoretical justification, that contains four parameters to be fitted. However, we will show that the nature of the discrepancy in $\alpha$ has a numerical origin. In other words, we will show that how the $\mathbf{q}\rightarrow 0$ and $\mathbf{G}=0$ limit is handled, is fully determining the values of the exciton binding energies, and hence the value of any empirical parameter $\alpha$ or function needed for the kernels.

In the optical limit and for $\mathbf{G}=0$, the matrix elements of Eq. (\ref{eq:xc-tddft}) have an indeterminate form $0/0$ for $f_{xc}^{\text{LRC}}$ type kernels, since the ground state KS valence and conduction states are orthogonal to each other. The usual way to handle this indetermination is to perform a series expansion of the numerator: %\cite{Yang2013, Byun2017} 
\begin{align}
    \bra{c\mathbf{k}} e^{i\mathbf{q}\cdot\mathbf{r}} \ket{v\mathbf{k}} \simeq \mathbf{q}\bra{c\mathbf{k}}\mathbf{r}\ket{v\mathbf{k}}.
    \label{pr}
\end{align}
which automatically entails a mutual cancellation of the vanishing $\mathbf{q}$ vectors in the numerator and denominator, and is a finite term. 
Nevertheless, a new problem arises as the interband transition value of the position operator has to be calculated, (right hand side of Eq [\ref{pr}]), being an ill-defined quantity (origin dependent) in infinite crystals with periodic boundary conditions. \\  

\begin{comment}
In practice, making use of the commutation relation  
\begin{align}
    \mathbf{\hat{p}}=i \comm{H_0}{\mathbf{\hat{r}}}_- \label{comm}, 
\end{align}
and the equality 
\begin{align}
\bra{2}\mathbf{\hat{p}}\ket{1}=i(\varepsilon_2 - \varepsilon_1)\bra{2}\mathbf{r}\ket{1}, \label{p-r}
\end{align}
%the response function and the $F^{\text{Hxc}}$ matrix elements are defined in the following way (for $\mathbf{G}= 0$): 
the matrix elements for $\mathbf{G}= 0$ are usually rewritten as  
\begin{align}
    F^{\text{Hxc},0}_{vc\mathbf{k}, v'c'\mathbf{k}'}=\frac{-2\alpha}{\mathcal{V}}
    &\frac{\bra{c\mathbf{k}}\mathbf{\hat{p}}+i\comm{V_{\text{NL}}}{\mathbf{\hat{r}}}\ket{v\mathbf{k}}}{E_{c\mathbf{k}} -E_{v\mathbf{k}}} \nonumber \\
    \times&\frac{\bra{c'\mathbf{k}'}\mathbf{\hat{p}}+i\comm{V_{\text{NL}}}{\mathbf{\hat{r}}}\ket{v'\mathbf{k}'}}{E_{c'\mathbf{k}'} -E_{v'\mathbf{k}'}} \label{FHxc_Ullr.}, 
\end{align}
where $V_{NL}$ is the non-local part of the potential\cite{}.*  \\ 

\end{comment}

In practice,  this problem is solved by using the commutator relation $\mathbf{\hat{p}}=i \comm{H_{\text{SCF}}}{\mathbf{\hat{r}}}$ that transforms the expected value of the position operator into a well defined expectation value of the momentum operator \cite{Resta-Position-operator,PhysRevB.33.7017}. This is commonly known as the \textit{p-r} relation: 
\begin{align}
\bra{c\mathbf{k}}\mathbf{r}\ket{v\mathbf{k}}=\frac{\bra{c\mathbf{k}}\comm{H_{\text{SCF}}}{\mathbf{\hat{r}}}\ket{v\mathbf{k}}}{(\epsilon_{c\mathbf{k}} -\epsilon_{v\mathbf{k}})}=
   \frac{ \bra{c\mathbf{k}}\mathbf{p}-i \comm{\mathbf{\hat{r}}}{V_{\text{nl}}} \ket{v\mathbf{k}}  }{(\epsilon_{c\mathbf{k}} -\epsilon_{v\mathbf{k}})}\label{p-r}.
\end{align} 
However, to derive Eq. (\ref{p-r}), is necessary to employ the above mentioned commutator relation which holds at each point in $\mathbf{r}$ space and to invoke the Hermiticity of the hamiltonian by $\bra{c\mathbf{k}}H_{\text{SCF}}=\bra{c\mathbf{k}}\epsilon_{c\mathbf{k}}$. In more algebraic detail, an integration by parts is carried out in the matrix element $\bra{c\mathbf{k}}H_{\text{SCF}}\mathbf{r}\ket{v\mathbf{k}}$ and a surface integral term $(\mathbf{C}_{c\mathbf{k},v\mathbf{k}})$ arises that should only be neglected for finite systems where the wave-function decays to zero far enough at the surface boundary \cite{Gu2013}. In contrast, we have for infinite solids with periodic boundary conditions that\footnote{The term $ \comm{\mathbf{\hat{r}}}{V_{\text{nl}}}$ has been omitted for notational simplicity as it plays no role in the argument.}: 
\begin{align}
    \bra{c\mathbf{k}}\mathbf{\hat{p}}\ket{v\mathbf{k}}=i(\epsilon_{c\mathbf{k}} -\epsilon_{v\mathbf{k}})\bra{c\mathbf{k}}\mathbf{\hat{r}}\ket{v\mathbf{k}}+\mathbf{C}_{c\mathbf{k},v\mathbf{k}} \label{r-C_terms}
\end{align}
with the additional surface term
\begin{align}
\begin{split}
    \mathbf{C}_{c\mathbf{k}',v\mathbf{k}}=\int_{S} \frac{d\mathbf{S}}{2}\cdot [\varphi^*_{c\mathbf{k}'}(\mathbf{r})\mathbf{\hat{p}}\varphi_{v\mathbf{k}}(\mathbf{r})+(\mathbf{\hat{p}}\varphi_{c\mathbf{k}'}(\mathbf{r}))^*\varphi_{v\mathbf{k}}(\mathbf{r})]\mathbf{r}.
\end{split}
\end{align}
Namely, the correct $\textit{p-r}$ relation contains a surface term that compensates for the ambiguity related to the position matrix element that depends on the choice of the unit cell. As the momentum matrix element does not have such an ambiguity, there must be an additional term. For an exhaustive review of the \textit{p-r} relation usage throughout the literature (sometimes incorrect) and the particular importance of this surface term, we invite the reader to check reference \cite{Gu2013}. In the present paper, we have first reproduced the same quantification of the term $\mathbf{C}_{c\mathbf{k},v\mathbf{k}}$ performed by the authors for GaAs, and we have included another two materials of our interest: GaN semiconductor and MgO insulator. We will see below the effect that it produces in the binding energies of excitons. \\

Using a simple Cohen-Bergstresser pseudopotential approach\cite{Cohen1966} it is possible to calculate the band structure and wave functions, and hence, compute the interband position matrix element $\bra{c\mathbf{k}} \mathbf{r} \ket{v\mathbf{k}}$ and the correction term $\mathbf{C}_{c\mathbf{k},v\mathbf{k}}$ which are shown In Fig. \ref{fig:2}. The two magnitudes depend on the location of the cell over which the integrals are taken. 
We also find that the magnitude of the $\mathbf{r}$ matrix element is much smaller than that of the correction term. 
In the case of GaAs the mean value of the correction term is about 31 times larger than the position term, whereas in the case of $\beta$-GaN and MgO the difference rate is about 28 and 13, respectively.   
These results show that in periodic systems, the contribution of the correction term is far from trivial. 
When neglecting the correction term, the  $\bra{c\mathbf{k}} \mathbf{r} \ket{v\mathbf{k}}$ matrix elements are infrastimated, so that smaller $\alpha$ parameters are needed to reproduce experimental binding energies.  
%All this issues point to the need of going beyond pure-TDDFT to consistently calculate exciton binding energies in solids. 
Because of this dependence of results on the unit-cell, consistent calculations of the binding energies in solids must consider going beyond pure-TDDFT. \\

The next step is to determine the particular contribution of the \textbf{q}=0, \textbf{G}=0 Coulomb singular term in the calculation of exciton binding energies. We have then built our own code that evaluates the matrix elements of Eq. [\ref{eq:xc-tddft}] and solves the eigenvalue problem of Eq.[\ref{casida_eigen}] within the TDA approximation. As input for the ground-state, we use  the Kohn-Sham wave-functions and energies obtained by the QUANTUM ESPRESSO code \cite{espresso}. Furthermore, experimental lattice parameters and norm-conserving LDA pseudopotentials without any scissors or GW corrections were used to compare to reference values \cite{Yang2013,Byun2017}. A $20\times20\times20$ $\Gamma$-centered $\mathbf{k}$-point mesh, 3 valence bands and 5 conduction bands were used for GaAs. These parameters for the rest of the materials are: $16\times16\times16$, 3, 6 for $\beta$-GaN and MgO; $16\times16\times8$, 6, 9 $\alpha$-GaN and AlN; and  $8\times8\times8$, 3, 24 for LiF, Ar and Ne. 
The only difference between the present work and the calculation presented in reference \cite{Byun2017} lies in how the singular term $\mathbf{q}=\mathbf{G}=0$ is treated numerically. 

\begin{figure}[htp]
    \centering  
    \includegraphics[scale=0.7]{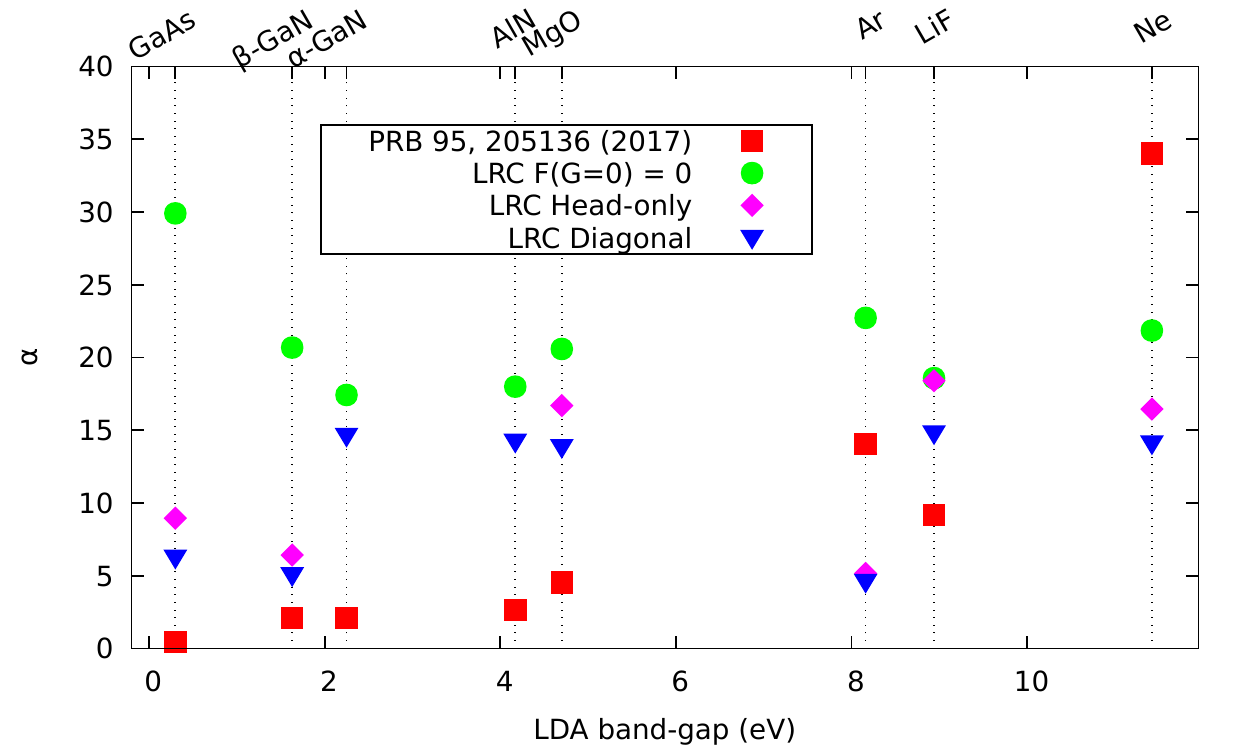}
    %\caption{Fitted LRC kernel $\alpha$-parameters that reproduce the experimental exciton binding energy. In red, values from reference\cite{Byun2017}. Values obtained omitting the singular term are given in green, while values obtained by treating the singular term by means of equation (\ref{singular_term}) are given in blue (diagonal kernel) and magenta (head-only kernel).   }
    \caption{Fitted LRC kernel $\alpha$-parameters that reproduce the experimental exciton binding energy.  Values obtained omitting the singular term are given in green, while the diagonal and head-only kernel are given in blue and magenta, respectively.  Red squares collects the values from reference \cite{Byun2017}.}
    \label{fig:1}
\end{figure}

We avoid to use the p-r relation trick of Eq[\ref{p-r}], and we have directly computed the matrix elements for a finite, but very small, $\mathbf{q}$ grid, as implemented in reference\cite{berkeleyGW}.  Figure \ref{fig:1} shows the obtained $\alpha$ values that brought the binding energies close to experiments for a set of semiconductors and insulators in the following three cases: a) setting $F_{xc}(\mathbf{G=0})=0$ in Eq.[\ref{eq:xc-tddft}], i.e., neglecting the indetermination; b) only considering the $F_{xc}(\mathbf{G=0})$ term different from zero and neglecting the all the other terms of the summatory (LRC Head-only); and c) full solution (LRC Diagonal). 
Moreover, for comparison purposes, Figure \ref{fig:1} includes the values for $\alpha$ obtained using the \textit{p-r relation}, taken from reference\cite{Byun2017}.  

The resulting $\alpha$ values for the (a) case, in which the indeterminate head-term is ignored, lay between 15 and 30, as shown by the green dots in Fig. \ref{fig:1}, and no noticeable trend with respect to the material type (semiconductor/insulator) is noticed. These values can be two orders of magnitude larger for small gap semiconductors and of the same order for large gap insulators when compared to those in ref \cite{Byun2017}. Thus, it necessarily implies that the $F_{xc}(\mathbf{G=0})$ contribution is crucial for completely fixing the final values of $\alpha$. In addition, when the head terms are added, either isolated (magenta diamonds) or including the complete summation in Eq. \ref{eq:xc-tddft} (blue triangles), the $\alpha$ values needed to reproduce the experimental binding energies decrease significantly. The  $\alpha$ values are still different from the red squares of the above mentioned reference. It is worth mentioning that in our opinion, the parabolic trend shown by red squares as the bandgap of the materials increase is somehow a consequence due to the use of the \textit{p-r relation} in which the surface term is ignored. Note that in Eq (\ref{p-r}) the bandgap energy divides each momentum matrix element, (and assuming that the numerator is bounded), higher values of $\alpha$ are required as we move towards insulating materials.

%%%%COMMENTS ON THE CORRECTION TERM Ccv 

\subsection{Hybrid TDDFT: Screened Exact Exchange kernel} 
One of the biggest computational bottlenecks of solving the Bethe-Salpeter Equation is building the screened Coulomb interaction matrix $W_{\mathbf{G,G}'}$ in equation (\ref{eq:xc-BSE})\cite{SXX-paper,Byun_Ullrich-kernels_2020}.
In the Screened Exact Exchange (SXX) approach, the diagonal terms of the exact exchange are considered and they are given by %\cite{SXX-paper}, 
\begin{equation}
    W_{\mathbf{G,G}'}(\mathbf{q}) = -4\pi\gamma\frac{\delta_{\textbf{GG}'}}{\abs{\mathbf{q+G}}^2}, 
    \label{eq:SXX-kernel}
\end{equation}
where the dielectric screening $\gamma$ is usually introduced as a parameter.  
A usual parametrization for this term is given by $\gamma = \epsilon^{-1}$,  
where $\epsilon$ is the experimental or $ab \ initio$ calculated static dielectric function \cite{SXX-paper, Sun-Yang-Ullrich-2020}.  
This approach has been shown to give comparable results to the full BSE\cite{SXX-paper,Byun_Ullrich-kernels_2020,Sun-Yang-Ullrich-2020}, while considerably lowering the computation time.

\subsubsection{Wigner-Seitz truncation of the kernel} %Ws truncation of the Coulomb potential 
%In the literature, it has been shown that the previously explained approach yields correct exciton binding energies\cite{}. This happens to be true even in the head-only SXX approach (h-SXX) proposed by Ullrich et al.\cite{}, where only the divergent $-4\pi\gamma\delta_{\textbf{GG}'}\delta_{\textbf{G}0}/\abs{\textbf{G}}^2$ term of the exchange is considered. \\ 

In the SXX approach, a correct treatment for the Coulomb term is crucial. However, in comparison with TDDFT, the $\mathbf{q=G=}0$ term for the $c=c'$ and $v=v'$ matrix elements now diverges as $1/0^2$. 
The use of real-space truncation methods to  handle numerically the Coulomb interaction have been previously studied to calculate the exact exchange energy in solids\cite{kim-PRL, kim-PRB, Sundararaman-Arias-2013}. 
They found that the real space truncation of the Coulomb term on a Wigner-Seitz shape supercell that naturally includes the crystal symmetry, is the best choice in terms of fast convergence of the k-point sampling. In this approach, the coulomb kernel is given as follows: 
\begin{equation}
    W_{\textbf{G}}^{\text{WS}} = \frac{4\pi}{G^2}\left(1 - e^{\frac{-G^2}{4\alpha^2}}\right) + \frac{\Omega}{N_{\textbf{r}}}\sum_{r \in \text{WS}}e^{-i\textbf{q}\cdot\textbf{r}}\frac{erf(\alpha r)}{r}, 
    \label{eq:K-WS}
\end{equation} 
which has a finite, well-behaved value at \textbf{G}=0:
\begin{equation}
    \bar{K}_{\textbf{G}=0}^{\text{WS}} = \frac{\pi}{\alpha^2} +  \frac{\Omega}{N_{\textbf{r}}}\sum_{r \in \text{WS}} \frac{erf(\alpha r)}{r},
\end{equation}
 where $erf$ is the error function, and $\alpha$ is the range-separation parameter. 

We have implemented this particular truncation method for the calculation of the matrix elements in Eq (\ref{eq:SXX-kernel}) and studied its performance for a testing set of semiconductors and insulators. We 
calculate the exciton binding energy with respect to the following convergence parameters: k-point sampling of the Brillouin Zone and dielectric screening parameter $\gamma$. 
%As a general trend, we 
We find that other convergence parameters, such as the number of valence and conduction bands, have a negligible impact on the calculated exciton binding energies. 
In other words, for all the compounds under analysis, a single valence and conduction band were enough to yield converged results. \\

\begin{figure}[htp]
    \centering
    \includegraphics[scale=0.30]{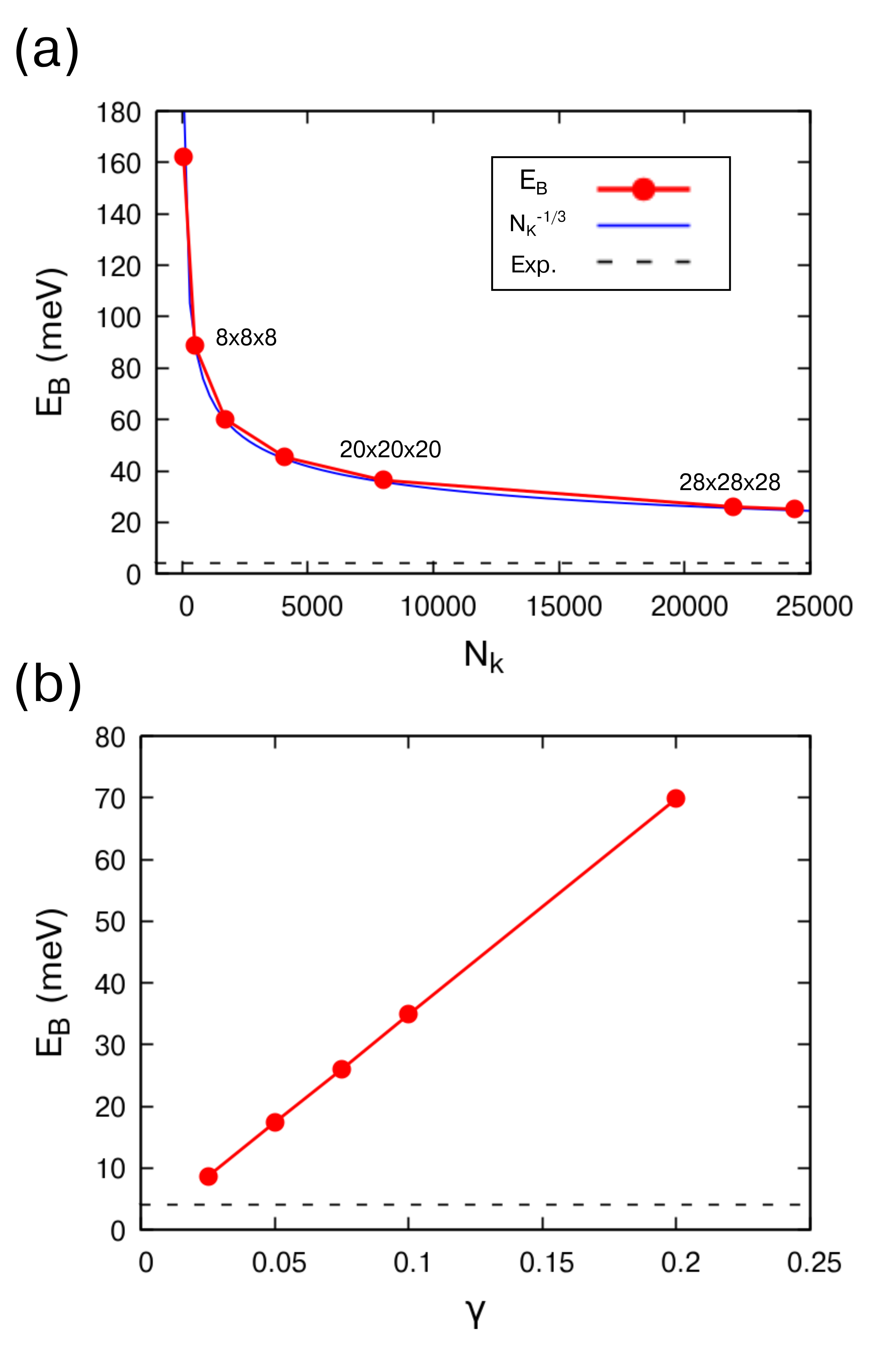}
    \caption{Calculated exciton binding energies of GaAs with respect to (a) the Brillouin Zone sampling $N_k$  and (b) the dielectric screening parameter $\gamma$.  For the k-point sampling, the exciton binding energy is proportional to $N_k^{-1}$, a trend that has already been observed in the literature.  For the dielectric screening, the binding energies vary linearly with $\gamma$. }
    \label{fig:3}
\end{figure}

%K-POINTS
 The convergence of the exciton binding energy with respect to the Brillouin Zone sampling (panel (a)) and dielectric screening parameter $\gamma$ (panel (b)) is shown in panels (a) and (b) of Fig. (\ref{fig:3}), respectively. 
We find that the exciton binding energy decays as $N_k^{-1/3}$, where $N_k$ is both the number of k-points in the Brillouin Zone and the number of unit cells in the WS supercell of the crystal in the real space. This slow convergence requires large k-point meshes, and the numerical results are highly dependent on $N_k$.  
%GAMMA 
For all the compounds under analysis, the binding energy is linearly proportional to  the dielectric screening $\gamma$. To keep all the calculation \textit{ab-initio}, our $\gamma$ values were computed using the RPA dielectric constant. Using the experimental dielectric constant changed the binding energies less than 10\% in the the wide gap insulators, and up to 30\% for GaAs and CdS.  
We therefore conclude that, even if the dielectric screening $\gamma$ is an important parameter to precisely determine the exciton binding energy, $N_k$ is the key variable. 
\\ 
%We remark that these trends are uniform*/identical for all the materials under analysis, without significant qualitative differences for insulators and semiconductors.  

%COMPARISON TO REFERENCE 
Table (\ref{tab:table2}) collects exciton binding energy values obtained in the WS-SXX approximation together with reference and experimental values. 
Our results can comparable with the literature, and we set the parameters to match those in the reference\cite{Sun-Yang-Ullrich-2020}, with the exception of the number of conduction and valence bands, which in our case were fixed to $n_c=n_v=1$. We find that for most of the semiconductors, the exciton binding energies are in good agreement with the theoretical results reported in the reference. Specifically, the Wigner-Seitz truncated kernel seems to perform better for Wurtzite type materials, i.e. w-GaN and AlN, because the numerical truncation of the $1/r$ potential on the WS supercell, takes into account more efficiently the elongated $c$ axis of the hexagonal lattice symmetry.  However, in the case of the insulators, we observe a clear underestimation of the binding energies with the WS-SXX kernel. We attribute the underestimation to the fact that the WS-SXX kernel mostly takes the long-range interaction into account, without proper consideration of the local field effects. In the wide gap insulators with localized Frenkel type excitons, these effects are not negligible, and this discrepancy might merit further investigation with fully hybrid kernels beyond the scope of this paper. 
Another explanation to the calculated values is that for all the compounds we observe identical trends with respect to the convergence parameters, with no qualitative variations with the electronic character. 
This results in exciton binding energies that are mostly governed by the chosen k-grid and $\gamma$ parameters. 
The sharp dependence on the k-grid is specially worrying, since fully converged results could vary greatly with the numerical values that are obtained from the grids that can be currently handled. All in all, we find that the sharp dependence on the singular term %, and therefore on the k-point sampling, 
presents an important numerical bottleneck for the calculation of precise exciton binding energies in solids, which limits the real applicability of the simple %xc 
kernels that are currently used.

%We compute the exciton binding energy of an array of semiconductors and insulators, and analyze their dependence on the following convergence parameters: k-point sampling of the Brillouin Zone, number of valence and conduction bands, screening parameter $\gamma$ and head-only versus full-SXX approaches. Figures () to () display the convergence of Eb with respect to these parameters, and the converged exciton binding energies are shown in Table () along the experimental and reference [] values. 

%We find the k-point sampling and the screening parameter to be the critical convergence parameters, while increasing the number of bands and using the full-SXX approach hardly vary the obtained results (see figures S() to S()). For all the compounds under analysis, the exciton binding energy is linearly proportional to gamma (see Fig. ()), and using the experimental and RPA dielectric constants produced a change of about 10*%.  
%In Fig. (), the convergence of the Eb of GaAs is shown with respect to the k-point grid employed, (...) 

%[Paragraph about the table; our Eb values compared to experiment/Ullrich. General trends, semiconductor/insulator difference... General thoughts about the method, posible improvements?? ] \\ 

%\begin{comment}
\begin{table}%[htp!]
\centering
\begin{tabular}{lllllllll}
\hline 
\hline 
Approach  & GaAs & CdS & z-GaN & w-GaN & AlN & MgO & LiF  & Ar   \\ 
\hline
BSE\cite{Sun-Yang-Ullrich-2020}                 & 24   & 59  & 103 & 110 & 181 & - & 2050 & 1830 \\ 
SXX\cite{Sun-Yang-Ullrich-2020}                 & 24   & 58  & 101 & 106 & 177 & - & 1930 & 1750 \\
%WS-SXX              &  25.3  & 57.6  & 91.9 & 95 & 124 & 180 & 588 & 387 \\ 
WS-SXX              &  25  & 58  & 92 & 95 & 124 & 180 & 588 & 387 \\ 
Exp.                & 4    & 28  &  26 &  20 &  75 & 80 & 1600 & 1900 \\ 
\hline 
\hline 
\end{tabular}
\caption{\label{tab:table2}%
Exciton binding energies obtained with the different hybrid approaches. 
The experimental data was extracted from Refs.\cite{GaAs-exp-1,GaN-AlN,z-GaN-exp,z-GaN-exp-2,CdS-exp,CdS-exp-2,GaN-exp,GaAs-exp-2,AlN-exp,Ar-exp,MgO-exp}. 
}
\end{table}
%\end{comment}

%%%%%%%%%%%%%%%%%%%%%%%%%%%%%%%%%%%%%%%%%%%%%%%%%%%%%%%%%%%% 
%%%%%%%%%%%%%%%%%%%%%%  CONCLUSIONS %%%%%%%%%%%%%%%%%%%%%%%%
%%%%%%%%%%%%%%%%%%%%%%%%%%%%%%%%%%%%%%%%%%%%%%%%%%%%%%%%%%%% 
\section{CONCLUSIONS}
In this paper, we studied the effect of the long-range Coulomb term in the calculation of the exciton binding energies. 
We focused our analysis on two different frameworks: LRC-TDDFT, and hybrid calculations with the SXX kernel. 
We find that in the pure TDDFT calculations, the effect of the correction term $C_{cv}$ is crucial, and that neglecting it could lead to errors. 
In the hybrid TDDFT-BSE framework, we propose a new WS truncated SXX kernel, which yields results close to the BSE for semiconductors, while falling short for the wide-band insulators. 
We once again find that the effect of the long-range Coulomb interaction is the leading contribution to the calculated exciton binding energies, concluding that a correct and careful description of this term is essential in this kind of calculation. 
We believe that the findings of this work encourage further and more detailed analysis of the Coulomb singularity that governs the excitonic effects in solids.

%%%%%%%%%%%%%%%%%%%%%%%%%%%%%%%%%%%%%%%%%%%%%%%%%%%%%%%%%%%%%%%%%%%% 
%%%%%%%%%%%%%%%%%%%%%%%%%%% APPENDIX %%%%%%%%%%%%%%%%%%%%%%%%%%%%%%% 
%%%%%%%%%%%%%%%%%%%%%%%%%%%%%%%%%%%%%%%%%%%%%%%%%%%%%%%%%%%%%%%%%%%%  
\appendix

\bibliography{exciton.bib}

\end{document}